\documentclass[showpacs,preprintnumbers,amsmath,aps,nofootinbib,amssymb,superscriptaddress]{revtex4}
\usepackage{graphicx}
\usepackage[english]{babel}
\usepackage{bm}
\usepackage[font=scriptsize]{caption}
\usepackage{ragged2e}
\usepackage[font=small,labelfont=bf,justification=raggedright]{caption}
\usepackage[utf8]{inputenc}
\usepackage[tikz]{bclogo}
\usepackage[framemethod=tikz]{mdframed}

\newsavebox{\measurebox}

\begin{document}
\baselineskip 16pt
\title{Hamilton's approach in cosmological inflation with an exponential potential and its observational constraints}
\author{Omar E. N\'u\~nez}
\email{neophy@fisica.ugto.mx}
\author{J. Socorro}
\email{socorro@fisica.ugto.mx}
\affiliation{Departamento de  F\'{\i}sica, DCeI, Universidad de Guanajuato-Campus Le\'on, C.P. 37150, Le\'on, Guanajuato, M\'exico}
\author{Rafael Hern\'{a}ndez-Jim\'{e}nez}
\email{s1367850@sms.ed.ac.uk}
\affiliation{School of Physics and Astronomy, University of Edinburgh, Edinburgh, EH9 3FD, United Kingdom}

\begin{abstract}
The Friedmann-Robertson-Walker (FRW) cosmology is analyzed with a general potential $\rm V(\phi)$ in the scalar field inflation scenario. 
The Bohmian approach (a WKB-like formalism) was employed in order to constraint a generic form of potential to the most suited to drive inflation, from here a family of potentials emerges; in particular we select an exponential potential as the first non trivial case and remains the object of interest of this work. The solution to the Wheeler-DeWitt (WDW) equation is also obtained for the selected potential in this scheme. Using Hamilton's approach and equations of motion for a scalar field $\rm \phi$ with standard kinetic energy, we find the exact solutions to the complete set of Einstein-Klein-Gordon (EKG) equations without the need of the slow-roll approximation (SR). In order to contrast this model with observational data (Planck 2018 results), the inflationary observables: the tensor-to-scalar ratio and the scalar spectral index are derived in our proper time, and then evaluated under the proper condition such as the number of e-folding corresponds exactly at 50-60 before inflation ends. The employed method exhibits a remarkable simplicity with rather interesting applications in the near future.

\end{abstract}


 \maketitle

\section{Introduction}
The inflation paradigm is considered the most accepted mechanism to explain many of the fundamental problems of the early stages in the 
evolution of our universe \cite{guth1981, linde1982, turner1981, starobinsky1980}, such as the flatness, homogeneity and isotropy observed 
in the present universe. Another important aspect of inflation is its ability to correlate cosmological scales that would otherwise be 
disconnected. Fluctuations generated during this early phase of inflation yield a primordial spectrum of density perturbation \cite{Starobinsky:1979ty,Mukhanov:1981xt,kodama, bassett}, 
which is nearly scale invariant, adiabatic and Gaussian, which is in agreement with cosmological observations \cite{Planck}. 
The single-field scalar models have been broadly used to describe such expansion, the most phenomenological successful are those with a 
quintessence scalar field and slow-roll inflation \cite{barrow, andrew1998b, ferreira2, copeland1, copeland2, andrew2007, gomez, capone, kolb}.

Research on the inflationary topic is primarily done in two forms, one of them is to modify the General Relativity in a way that allows 
the inflationary solutions. The other one is the introduction of new forms of matter, with the capability of driving inflation into the 
General Relativity, where one introduces a canonical scalar field. Essentially, in the studies of inflationary cosmology one imposes the 
usual slow roll approximation (SR) with the objective to extract expressions for basic observables, such as the scalar spectral index and the 
tensor-to-scalar ratio. The slow roll approximations reduce the set of Einstein-Klein-Gordon (EKG) equations in such a way that one can quickly 
obtain the solution to the scale factor in this approximation. However, there is also an alternative approach which allows for an easier
derivation of many inflation results without such approximation, and that is known as the \emph{Hamilton's formulation}, which is widely 
used in analytic mechanics. Using this approach we obtain the exact solution of the complete set of EKG equations without 
using the aforementioned approximation.

There are many works in the literature that have extensively treated this type of problems, such as \cite{Lucchin:1984yf,Salopek:1990jq,ratra1, ratra2,russo,Andrianov:2011fg,Piedipalumbo:2011bj,Fre:2013vza}, where the authors obtained exact and SR dynamical solutions of a system such as the scalar field potential employed in the form $V(\phi)\propto e^{-\frac{1}{\sqrt{p}}\phi}$. 
Prominently in \cite{Lucchin:1984yf} a radiation fluid was included and an extensive study of the evolution of perturbation in power-law inflation was performed. And a thoroughly classical and quantum analysis, considering dynamical and perturbative aspects, was implemented in \cite{Salopek:1990jq,ratra1, ratra2,russo,Andrianov:2011fg,Piedipalumbo:2011bj,Fre:2013vza}.
However, even if the above mentioned works rigorously examined the scalar field fluctuations, the observational parameters were not contrasted with the up today most accurate astronomical surveillance data, therefore one of the main reasons for this analysis. Further ahead the results are discussed.

Even more, in \citep{russo} the author deals with different values of $\rm \lambda$ in a $\rm V(\phi)\propto e^{-\lambda\phi}$ like potential as a model for the primordial inflation. Indeed, Russo's model is similar to ours albeit treated with a different approach. Russo's solution's are equivalent as well, we were even able to find a direct transformation for the critical value of $\rm \lambda=\sqrt{3}$, where he finds the following set of solutions, $\rm A(\tau)=e^{\frac{1}{3}\tau^2}(2\tau^{\frac{1}{6}})$ 
and $\rm \phi(\tau)=\frac{1}{\sqrt{3}}(2\tau^2-\ln 2\tau)$, we can see that under a particular transformation we can change our solutions to his
proper time $\tau$, which is $\rm \tau=e^{12\sqrt{3}P_0 t}$, for this particular case, the solutions we obtain with our method are $$\rm \phi(t)=\phi_0-12P_0 t+ \frac{\tilde{P}}{6P_0}e^{24\sqrt{3}P_0 t} \, , \hspace{1cm} \rm A(t)=A_0 Exp{\left[\frac{\tilde{P}}{12\sqrt{3}P_0}e^{24\sqrt{3}P_0 t}+2\sqrt{3}P_0 t\right]}$$ where $\rm (\tilde{P}, P_0)$ are integration constants, substituting $\rm \tau$ in such solutions yields similar results. However, for the particular case of $\rm \lambda<1$, Russo's analysis concludes that the obtained inflation is eternal, but we found otherwise, we were able to compute the observables at the appropriate number of e-folding that ensures the end of inflation and compare it with the observational data; thus, imposing a more rigorous analysis of the solutions, such results are discussed within the work.

There are other works of interest that have treated with a similar model, and even under the premise of Super Symmetric Quantum Mechanics \cite{sodo,ssw,sdp} where similar solutions are found.

In our approach we use the quantum solution of the Wheeler-DeWitt (WDW) equation, a basic equation in quantum cosmology, from where we 
obtain a family of scalar potentials in the Bohmian formalism as in previous works \cite{wssa,nuevo}, which are found using a constraint equation on the superpotential 
function $\rm S$ and the amplitude of probability $\rm W$, which are the solution to the Hamilton-Jacobi equation at quantum level, from here we select an exponential potential for the scalar field as the first non trivial case, which becomes the model of interest for this work. Thus, in that sense, we analyze the case of scalar field cosmology, constructed using a scalar field, and a general potential of the form $\rm V(\phi)=V_0 e^{-\lambda \phi}$, for different values of $\rm \lambda$ for which we find the 
exact solutions to the EKG equations. 

In order to thoroughly contrast the model we analyze the inflationary observables such as the number of e-folds, the tensor-to-scalar ratio and the primordial tilt or scalar spectral index. The analysis of the observables was performed in the same framework as the Planck Collaboration (2018) \cite{Planck}. The acceleration parameter was computed and used as a constraint on the solutions so that only those that result in a positive acceleration are considered. The analysis was performed using the scalar-field velocity as the parameter of evolution. Finally the results and discussions are presented in the last section of this work.
   
\bigskip

This work is arranged as follows. In section \ref{model} we present the model with the action and the corresponding EKG equations for our cosmological 
model under consideration and the Hamiltonain density as well. In section \ref{qap} we use the Hamiltonian density to compute the corresponding WDW equation, which is solved by using the Bohmian approach (a WKB-like formalism), from there we obtain a family of potentials which are suited to model inflation; we also obtain the exact solution to the WDW equation 
using the separation variable method with a general scalar potential. In section \ref{classical} a complete and exact classical representation of a canonical scalar field with exponential potential in a flat FRW metric is presented; in subsection \ref{obser} the inflationary observables are derived and the conditions such as the e-folding function $\rm N_e$ and the acceleration parameter are also computed and imposed on the solutions, the results are presented as well. Finally, in section \ref{conclusions} we present our conclusions for this work.

\section{The model \label{model}}

We begin with the construction of the scalar field cosmological paradigm, which requires a canonical scalar field $\rm \phi$. The action of a universe with the constitution of such field is

\begin{equation}
\rm {\cal L}[g,\phi]= \sqrt{-g} \left(\frac{M_{P}^{2}}{2}R-\frac{1}{2}g^{\mu\nu}\nabla_{\mu}\phi \nabla_{\nu}\phi+V(\phi) \right) \,,\label{lagra}
\end{equation}
where $\rm R$ is the Ricci scalar, $\rm V(\phi)$ is the corresponding scalar field potential, and $M_{P}^{2}=1/8\pi G$ denotes the reduced Planck mass. The corresponding variation of Eq.(\ref{lagra}), with respect to the metric and the scalar field gives the Einstein-Klein-Gordon field equations
\begin{eqnarray}
\rm R_{\alpha\beta}-\frac{1}{2}g_{\alpha\beta}R &=& -\frac{1}{2}\left(\nabla_\alpha\phi\nabla_\beta\phi-\frac{1}{2}g_{\alpha\beta}g_{\mu\nu}\nabla_\mu\phi\nabla_\nu\phi\right)+\frac{1}{2}g_{\alpha\beta}V(\phi) \,, \label{camrel} \\
\rm \square\phi-\frac{\partial V}{\partial\phi} &=& g^{\mu\nu} \phi_{,\mu\nu} -g^{\alpha \beta} \Gamma^\nu_{\alpha
\beta} \nabla_\nu \phi -\frac{\partial V}{\partial\phi}=0\,.\label{klein}
\end{eqnarray}
From  Eq.(\ref{camrel}) it  can be deduced  that the energy-momentum tensor associated with the scalar field is
\begin{equation}
 \frac{T^{(\phi)}_{\alpha\beta}}{M_{P}^{2}}=\frac{1}{2}\left(\nabla_\alpha\phi\nabla_\beta\phi-\frac{1}{2}g_{\alpha\beta}
g_{\mu\nu}\nabla_\mu\phi\nabla_\nu\phi\right)-\frac{1}{2}g_{\alpha\beta}V(\phi) \,.
\end{equation}\\
 The line element to be considered in this work is the flat FRW
\begin{equation}
\rm ds^2=-N(t)^2 dt^2 +e^{2\Omega(t)} \left[dr^2
+r^2(d\theta^2+sin^2\theta d\phi^2) \right], \label{frw}
\end{equation}
where $\rm N$ is the lapse function, which in a special gauge one can directly recover the cosmic time $\rm t_{phys}$ ($\rm Ndt=dt_{phys}$), the scale factor $\rm A(t)=e^{\Omega(t)}$ is in the Misner's parametrization, and the scalar function has an interval, $\rm \Omega \in (-\infty,\infty)$. The classical solution to Einstein-Klein-Gordon Eqs.(\ref{camrel},\,\ref{klein}) can be found using the Hamilton's approach, so we need to build the corresponding Lagrangian and Hamiltonian densities for this cosmological model. In that sense, we use Eq.(\ref{frw}) into Eq.(\ref{lagra}) having 

\begin{equation}
\rm {\cal L}=e^{3\Omega}\left[6\frac{\dot{\Omega}^{2}}{N}-\frac{1}{2M_{P}^{2}}\frac{\dot{\phi}^{2}}{N}+\frac{1}{M_{P}^{2}}NV(\phi)\right] \,,
\label{lagrafrw}
\end{equation}
the momenta and the associated velocities are

\begin{equation}
\rm \Pi_{\Omega}=\frac{12e^{3\Omega}}{N}\dot{\Omega}\,, \quad\quad \dot{\Omega}=\frac{N}{12}e^{-3\Omega} \Pi_{\Omega}\,,
\end{equation}
and
\begin{equation}
\rm \Pi_{\phi}=-\frac{e^{3\Omega}}{M_{P}^{2} N}\dot{\phi}\,, \quad\quad \dot{\phi}=-M_{P}^{2}Ne^{-3\Omega}\Pi_{\phi}\,. 
\end{equation}

The Hamiltonian density, written as $\rm {\cal L}=\Pi_q \dot q-N{\cal H}$, when performing the variation of this canonical Lagrangian with respect to $\rm N$, i.e. $\rm \delta {\cal L}/\delta N =0$, results that it is weakly zero: $\rm {\cal H}=0$. Hence the Hamiltonian density is

\begin{equation}\label{hamifrw}
\rm {\cal H}=\frac{e^{-3\Omega}}{24}\left[\Pi_{\Omega}^{2}-12M_{P}^{2}\Pi_{\phi}^{2}-24\frac{V(\phi)}{M_{P}^{2}}e^{6\Omega}\right] \,.
\end{equation}

\section{Quantum Approach\label{qap}}

The Wheeler-DeWitt equation has been treated in many different ways and there are a lot of papers that deal with different approaches to solve it, for instance in \cite{Gibbons}, they asked the question of what a typical wave function for the universe is. In \cite{Zhi} there is an excellent summary of a paper on quantum cosmology where the problem of how the universe emerged from a big bang singularity can no longer be neglected in the GUT epoch. On the other hand, the best candidates for quantum solutions are those that have a damping behaviour with respect to the scale factor, since only such wave functions allow good classical solutions when using the WKB approximation 
for any scenario in the evolution of our universe \cite{HH,H}.\newline

The Wheeler-DeWitt equation for this model is acquired by replacing $\rm \Pi_{q^\mu}=-i\hbar \partial_{q^\mu}$ in Eq.(\ref {hamifrw}) {\footnote{Only for convenience in this section we set $M_{P}=1$}}. The factor $\rm e^{-3\Omega}$ may be factor ordered with $\rm \hat \Pi_\Omega$ in many ways. Hartle and Hawking \citep{HH} have suggested what might be called a semi-general factor ordering, which in this case would order $\rm e^{-3\Omega} \hat \Pi^2_\Omega$ as

\begin{eqnarray}
\rm - e^{-(3- Q)\Omega}\, \partial_\Omega e^{-Q\Omega}
\partial_\Omega&=&\rm - e^{-3\Omega}\, \partial^2_\Omega +
 Q\, e^{-3\Omega} \partial_\Omega \,, \label {hh}
\end{eqnarray}

where  $\rm Q$ is any real constant that measures the ambiguity in the factor ordering for the variable $\rm \Omega$. In the following we will assume such factor ordering for the Wheeler-DeWitt equation, which becomes

\begin{equation}
\rm \hbar^2 \Box \Psi+ \hbar^2 Q\frac{\partial \Psi}{\partial
\Omega}- e^{6\Omega}U(\varphi)\Psi=0 \,, \label{wdwmod}
\end{equation}
where the field was re-scaled as $\rm \phi=\sqrt{12}\varphi$, and $\rm \Box=-\frac{\partial^2}{\partial \Omega^2}+\frac{\partial^2}{\partial \varphi^2}$ is the d'Alambertian in the coordinates $\rm q^\mu=(\Omega,\varphi)$ and the potential is $\rm U=  +24V(\varphi) $.

\subsection{The Bohmian formalism}

The main idea of this approach is the use of a WKB-like ansatz for the wave function
\begin{equation}
\rm \Psi = W(\ell^\mu) e^{- \frac{S_\hbar}{\hbar}(\ell^\mu)} \,,
\label{ans}
\end{equation}
where $\rm S_\hbar(\ell^\mu)$ is known as the superpotential function and $\rm W$ is the amplitude of probability that is employed in Bohmian formalism \citep{bohm}, which is then introduced into Eq.(\ref{wdwmod}), obtaining
\begin{equation}\rm
 \hbar^2 \left[{\Box \, W} - \frac{1}{\hbar}W {\Box \, S_\hbar} - \frac{2}{\hbar} {\nabla W}\cdot {\nabla
 S_\hbar}+\frac{1}{\hbar^2} W  \left(\nabla S_\hbar\right)^2\right]
 +\hbar^2 Q\left[ \frac{\partial W}{\partial \Omega} -  \frac{1}{\hbar}W \frac{\partial
S_\hbar}{\partial \Omega}\right] - {\cal U} W = 0 \,, \label {mod}
\end{equation}
which are ordered in power of  $\rm \hbar$,
\begin{equation}
\rm \hbar^2\left[ {\Box \, W} + Q \frac{\partial W}{\partial
\Omega}\right] - \hbar \left[W {\Box \, S_\hbar} + 2 {\nabla W}\cdot
{\nabla  S_\hbar}+ Q W \frac{\partial S_\hbar}{\partial
\Omega}\right] + W\left[\left(\nabla S_\hbar \right)^2- {\cal U}
\right]=0 \,.
\end{equation}
From this expansion we can see that the first term corresponds to the quantum potential in a Hamilton-Jacobi like equation, for $\rm \hbar^2$ we have a constraint equation and for the linear $\rm\hbar$ we have the imaginary part, thus
\begin{subequations}
\label{WDWa}
\begin{eqnarray}
\rm (\nabla S_\hbar)^2 - {\cal U} &=& 0  \quad [Einstein-Hamilton-Jacobi] \,, \label{hj} \\
\rm    \Box \, W + Q \frac{\partial W}{\partial \Omega} & = & 0 \quad [constraint \,\, equation] \,, \label{cons}  \,\\
\rm   W \left( \Box S+ Q \frac{\partial S_\hbar}{\partial \Omega}
  \right) + 2 \nabla \, W \cdot \nabla \, S_\hbar &=& 0 \quad [imaginary \,\, part] \,. \label{wdwmo} 
\end{eqnarray}
\end{subequations}

\subsection{Our approach to obtain a family of scalar potentials}

The following approach has been proposed before (see \citep{wssa}), so we use the  same steps to obtain the corresponding scalar
potential. First we solve Eq. (\ref{hj})
\begin{equation} \rm -\left(\frac{\partial S}{\partial
\Omega} \right)^2 +\left(\frac{\partial S}{\partial \varphi}
\right)^2=e^{6\Omega} U(\varphi), \label{hj-n}
\end{equation}
and using the following ansatz for the superpotential function $\rm S=e^{3\Omega}g(\varphi)/\mu$, then Eq.(\ref{hj-n}) becomes an ordinary differential equation for the unknown function $\rm g(\varphi)$ in terms of the scalar potential $\rm U(\varphi)$,
\begin{equation}
\rm \left(\frac{dg}{d\varphi} \right)^2-9g^{2}=\mu^2 U(\varphi), \label{master}
\end{equation}
this equations has several exact solutions, which can be generated in the following way when we consider that
$\rm U(\varphi)=g^2(\varphi) G(g)$, 
where $\rm g(\varphi)$ and $\rm G(g)$ are functionals of the argument, yet to determine, then introducing this into Eq. (\ref{master}), this one can be written in quadrature as
\begin{equation}
\rm d\varphi=\rm \pm \frac{dg}{g\sqrt{9+ \mu^2 G}}, \label{varphi}
\end{equation}
in Table \ref{t:solutions} appears the family of potentials and its corresponding generic functions found using this approach for the inflationary epoch,
\begin{center}
\captionsetup{width=0.9\textwidth}
    \begin{table}[ht]
            \begin{tabular}{|c|c|c|}
            \hline
             $\rm G(g)$& $\rm g(\varphi)$  &$\rm V(\varphi)$   \\ \hline
               0& $\rm g_0 e^{\pm \sqrt{3}  \Delta \varphi}$ & 0  \\ \hline
            $\rm  G_0$ &$\rm g_0 e^{\pm \frac{\lambda}{2} \Delta \varphi}$& $\rm V_0 e^{\pm \lambda \Delta \varphi}$ \\ \hline
              $\rm  G_0 g^{2}  $ & $\rm g_0 csch \left[3\Delta \varphi \right] $ & $\rm V_0  csch^4 \left[3\Delta \varphi \right]$ \\ \hline
            $\rm  G_0 g^{-2}  $  & $\rm g_0 sinh \left[3\Delta \varphi \right] $ & $\rm V_0$\\ \hline
             $\rm G_0 g^{-n}$ ($\rm n \neq 2$)&  $\rm g_0\, \left[sinh^2\left(\frac{3n\Delta \varphi}{2}\right)\right]^{1/n}$&
             $\rm V_0\, \left[cosh^2\left(\frac{3n\Delta \varphi}{2}\right)-1\right]^{\frac{2-n}{n}}$   \\\hline
             $\rm  G_0\, \ln g$& $\rm e^{v(\varphi)}$ &$\rm V_0  v(\varphi)e^{2v(\varphi)} $ \\
           & $\rm v(\varphi)=\left(\frac{3}{2}\Delta \varphi \right)^2$& \\ \hline
            $\rm G_0  (\ln g)^2$ &$\rm e^{\omega(\varphi)} $& $\rm V_0  \omega^2(\varphi)e^{2\omega(\varphi)} $ \\
          &  $\rm \omega(\varphi)=sinh(3\Delta \varphi) $ & \\
          \hline
            \end{tabular}
             \caption{ \label{t:solutions} The exact solutions to Eq. (\ref{varphi}) are presented. Each row represents a different independent solution and each column represents the form that each generic function must become for that specific solution and its corresponding potential.}
        \end{table}
    \end{center}

To solve the equation for the W function, Eq.(\ref{wdwmo}), we introduce the ansatz $\rm W =e^{u(\Omega)+v(\varphi)}$, by using the separation variable method we obtain the following two equations for the functions (u,v),
\begin{eqnarray}
\rm 2 \frac{du}{d\Omega} - Q &=& \rm  3k, \label{u}\\
\rm  \frac{d^2g}{d\varphi^2}+ 2 \frac{dv}{d\varphi}\frac{dg}{d\varphi}&=& \rm 3(3+k)g\,, \label{v}
\end{eqnarray}
where we have implemented a separation constant as 3k for simplicity. The corresponding solutions become
\begin{eqnarray}
\rm u(\Omega)&=& \rm \frac{k+Q}{2}\Omega + u_0 \,,\\
\rm v(\varphi)&=& \rm \frac{3(3+k)}{2} \int \frac{d\varphi}{\partial_\varphi [ln(g)]} -\int \frac{\frac{d^2 g}{d\varphi^2}}{\partial_\varphi g} + v_0 \,,\\
&=& \rm \frac{3(3+k)}{2} \int \frac{d\varphi}{\partial_\varphi [ln(g)]} -\frac{\mu^2}{2}\int \frac{d[U(\varphi)]}{(\partial_\varphi g)^2} + v_0 \,,
\end{eqnarray}
and for the last equation, we use Eq. (\ref{master}) for this transformation. Then, the function W has the following form
\begin{equation}
\rm W= W_0 Exp\left[\frac{k}{2}\Omega + \frac{3(3+k)}{2} \int \frac{d\varphi}{\partial_\varphi [ln(g)]} \right]\, 
Exp \left[ \frac{Q}{2}\Omega -\frac{\mu^2}{2}\int \frac{d[U(\varphi)]}{(\partial_\varphi g)^2} \right], \label{w}
\end{equation}

If we restrict the solutions in Eqs. (\ref{master},\,\ref{w}) to comply with the constraint Eq. (\ref{cons}), this results in the next conditions
 \begin{equation}
 \rm \frac{d^2v}{d\varphi^2}+\left( \frac{dv}{d\varphi} \right)^2 - \frac{k^2- Q^2}{4}=0\,,\qquad \frac{dv}{d\varphi}=\frac{3(3+k)}{2} \frac{1}{\partial_\varphi [ln(g)]} -\frac{\mu^2}{2} 
\frac{\frac{d[U(\varphi)]}{d\varphi}}{(\partial_\varphi g)^2} \,.
 \end{equation}

Considering the particular case for the function $\rm g=e^{-\frac{\lambda}{2}\varphi}$, and its corresponding scalar potential is
$\rm U(\varphi)=V_0 e^{-\lambda \varphi}$, yields a constraint between all constants
\begin{equation}
\rm \left( \frac{2\mu^2 V_0 - 3(3+k)}{\lambda}\right)^2 -  \frac{k^2- Q^2}{4}=0 \,,
\end{equation}
thus, the amplitude of probability W becomes
\begin{equation}
\rm W=W_0 Exp \left\{\frac{k+Q}{2}\left[\Omega + \pm (k-Q)\varphi\right]  \right\} \,.
\end{equation}

However, to study all the obtained potentials is an exhaustive work and as such we chose the exponential potential as the first non trivial case to model inflation, and it remains as the object of interest through this work.

\section{Classical solutions using an exponential potential of the form: $\rm V=V_{0}e^{-\frac{\lambda\phi}{M_{P}}}$\label{classical}}

Using the gauge $\rm N=24e^{3\Omega}$, the Hamiltonian density is

\begin{equation}\label{hamil-class}
\rm {\cal H}=\Pi_{\Omega}^{2}-12M_{p}^{2}\Pi_{\phi}^{2}-24\frac{V_{0}}{M_{p}^{2}}e^{6\Omega-\frac{\lambda\phi}{M_{P}}} \,,
\end{equation}
working with the Hamilton's equations of motion we have the canonical velocities and momenta
\begin{eqnarray}
\rm \dot{\Omega} &=& 2\Pi_{\Omega} \,,  \qquad\qquad \dot{\Pi}_{\Omega} =  144\frac{V_{0}}{M_{p}^{2}}e^{6\Omega-\frac{\lambda\phi}{M_{P}}} \,,\\
\rm \dot{\phi} &=& -24M_{P}^{2}\Pi_{\phi}\,, \quad\,\,  \dot{\Pi}_{\phi} =  -24\lambda\frac{V_{0}}{M_{p}^{3}}e^{6\Omega-\frac{\lambda\phi}{M_{P}}} \,,
\end{eqnarray}
from here one can obtain a relation between $\rm \Pi_\Omega$ and $\Pi_\phi$ as 
\begin{equation}
\rm \frac{\dot{\Pi}_{\Omega}}{\dot{\Pi}_{\phi}}=-\frac{6M_{P}}{\lambda} \,,
\end{equation}
yielding a relation between the two momenta 

\begin{equation}\label{Piphi-PiOmega}
\rm \Pi_{\phi}=-\frac{\lambda}{6M_{P}}\Pi_{\Omega}+P_{0} \,,
\end{equation}
where $\rm P_{0}$ is an integration constant and remains a free parameter of the model to be adjusted with the observable data. Substituting Eq.(\ref{Piphi-PiOmega}) in the Hamiltonian density, Eq.(\ref{hamil-class}) ($\rm {\cal H}=0$), we arrive to the following relation

\begin{equation}
\rm \dot{\Pi}_{\Omega}+2\left(\lambda^{2}-3\right)\Pi_{\Omega}^{2}-24M_{P}\lambda P_{0}\Pi_{\Omega}+72M_{P}^{2}P_{0}^{2}=0
\end{equation}
which solution is 

\begin{equation}\label{general-solution-of-Pi-Omega}
\rm \frac{1}{24\sqrt{3}M_{P}P_{0}}\log\left[\frac{(\lambda^{2}-3)\Pi_{\Omega}-6(\lambda+\sqrt{3})M_{P}P_{0}}{(\lambda^{2}-3)\Pi_{\Omega}-6(\lambda-\sqrt{3})M_{P}P_{0}}\right]=-t+P_{1}\,,
\end{equation}
where $\rm P_1$ is a time-like integration constant. With Eq.(\ref{general-solution-of-Pi-Omega}), the canonical momentum $\Pi_{\phi}$ and the rest of variables can be solved. Hence the solutions are 

\begin{eqnarray}
\rm \Pi_{\Omega}(t) &=& \frac{6\lambda M_{P}P_{0}}{\lambda^{2}-3}+\frac{6\sqrt{3}M_{P}P_{0}}{\lambda^{2}-3}\coth\left[12\sqrt{3}M_{P}P_{0}(t-P_{1})\right] \,,\\
\rm \Pi_{\phi}(t) &=& -\frac{3P_{0}}{\lambda^{2}-3}-\frac{\sqrt{3}\lambda P_{0}}{\lambda^{2}-3}\coth\left[12\sqrt{3}M_{P}P_{0}(t-P_{1})\right] \,,\\
\rm \phi(t) &=& \phi_{0}+\frac{72M_{P}^{2}P_{0}t}{\lambda^{2}-3}+\frac{2\lambda M_{P}}{\lambda^{2}-3}\log\left[\sinh\left[12\sqrt{3}M_{P}P_{0}(t-P_{1})\right]\right] \label{phit}\,,\\
\rm \Omega(t) &=& \Omega_{0}+\frac{12\lambda M_{P}P_{0}t}{\lambda^{2}-3}+\frac{1}{\lambda^{2}-3}\log\left[\sinh\left[12\sqrt{3}M_{P}P_{0}(t-P_{1})\right]\right] \,.\label{Omegat}
\end{eqnarray}
This set of solutions is a complete and exact classical representation of a canonical scalar field with exponential potential in a flat FRW metric. Also from Eqs. (\ref{phit},\,\ref{Omegat}) we find the following relation between the coordinates fields $\rm (\Omega,\phi)$: $\rm \Delta \phi= 2\lambda M_P \Delta \Omega - 24M_P^2 P_0 t$. Moreover, $\rm V_{0}$ is

\begin{equation}
\rm V_{0}=-\frac{3}{2}\frac{P_{0}^{2}}{(\lambda^{2}-3)}\exp\left[\frac{\lambda\phi_{0}}{M_{P}}-6\Omega_{0}\right]M_{P}^{4}\,,
\end{equation}
for $\rm V_{0}>0$ we have that $\rm \lambda^{2}<3$; constricting the domain of $\rm \lambda$ we subsequently work with.

\subsection{Observables\label{obser}}

Inflation is characterised by the number of e-folds it expands during such period, that corresponds to $\rm A''_{phys}>0$, where the primes represent the derivatives with respect to the cosmic time $\rm t_{phys}$. The e-folding function $\rm N_{e}=\int dt_{phys}H(t_{phys})$ is described by $\rm t_{phys}$: computing the integral from $\rm t_{phys}*$ to $\rm t_{phys\,\,end}$; where $\rm t_{phys}*$ represents the time when the relevant cosmic microwave background (CMB) modes become superhorizon at 50-60 e-folds before inflation ends at $\rm t_{phys\,\,end}$; and $\rm H(t_{phys})=H_{phys}=A'_{phys}/A_{phys}$ is the Hubble parameter. Although, in our prescription we use a proper time $\rm t$, we can evaluate the Hubble function in the corresponding gauge as $\rm H_{phys}=\dot{\Omega}/N$. Moreover, in order to compute the function $\rm N_{e}$, it is convenient to use another variable to describe such quantity; let us use $\rm \dot{\phi}$, the scalar field velocity, instead of the proper time $\rm t$ as an evolution parameter. Hence, taking the time derivative of Eq.(\ref{phit}), we have 

\begin{equation}\label{t1}
\rm t=\frac{1}{12\sqrt{3}M_{P}P_{0}}\text{arccoth}(x)+P_{1}\,,
\end{equation}
where

\begin{equation}\label{x-dotphi}
\rm x= \frac{\lambda^{2}-3}{24\sqrt{3}\lambda M_{P}^{2}P_{0}}\dot{\phi}-\frac{\sqrt{3}}{\lambda} \,,
\end{equation}
this variable $\rm x$ must satisfy that $\rm |x|>1$ in order to have $\rm t\,\epsilon\,\Re$. At the end of inflation the expansion rate of the scale factor must be null which translates to $\rm -H'_{phys}=H_{phys}^{2}$ or $\rm\ddot{\Omega}=2\dot{\Omega}^{2}$. From here we can compute the time when inflation ends, which is

\begin{equation}
\rm t_{end}=\frac{1}{12\sqrt{3}M_{P}P_{0}}\text{arccoth}\left[\mp\frac{\sqrt{3}}{3}\left[\frac{\lambda\pm 3}{\lambda\pm 1}\right]\right]+P_{1}\,.
\end{equation}

Then comparing above result and Eq.(\ref{t1}), at the end of inflation $\rm \dot{\phi}_{end}$ is

\begin{equation}\label{dotphiend}
\rm \dot{\phi}_{end}=\pm \frac{24M_{P}^{2}P_{0}}{\lambda \mp 1} \,.
\end{equation}

Henceforth we will use $\rm \dot{\phi}$ as the evolution parameter. The number of e-folds is

\begin{equation}\label{number-of-e-fold}
\rm N_{e} = \frac{\sqrt{3}}{6(\lambda^{2}-3)}\left\{24\sqrt{3}\lambda M_{P}P_{0}P_{1} +(\sqrt{3}-\lambda)\log\left[\frac{x_{*}+1}{x_{end}+1}\right] +(\sqrt{3}+\lambda)\log\left[\frac{x_{*}-1}{x_{end}-1}\right]\right\} \,, 
\end{equation}
where $\rm x_{*}$, implying $\dot{\phi}_{*}$, is to be determined at 50-60 e-folds before inflation ends, and 

\begin{equation}\label{xend}
\rm x_{end}=-\frac{\sqrt{3}}{\lambda}\pm\frac{\sqrt{3}}{3\lambda}\left(\frac{\lambda^{2}-3}{\lambda \mp 1}\right) \,. 
\end{equation} 

For the standard Bunch-Davies vacuum phase space distribution at the time when observable CMB scales leave the horizon during inflation, $\rm t_{phys *}$, the scalar amplitude of the primordial perturbations $\rm \Delta_{\mathcal{R}}^2$ has the form \cite{bassett}

\begin{equation}\label{DeltaR-physical-time}
\rm \Delta_{\mathcal{R}}^{2}=\frac{H_{phys *}^{2}}{4\pi^{2}}\frac{H_{phys *}^{2}}{\phi'^{2}_{phys *}} \,,
\end{equation}
where $\rm H_{phys *}=\dot{\Omega}_{*}/N_{*}$ and $\rm \phi'_{phys *}=\dot{\phi}_{*}/N_{*}$; therefore in our proper time we have

\begin{equation}\label{DeltaR-proper-time}
\rm \Delta_{\mathcal{R}}^{2}=\frac{1}{4\pi^{2}}\frac{\dot{\Omega}_{*}^{4}}{N_{*}^{2}\dot{\phi}_{*}^{2}}\,.
\end{equation}

We need to have a relation of $\rm \Omega_{*}=\Omega(\dot{\phi}_{*})$ (given that $\rm N=24e^{3\Omega}$) and $\rm \dot{\Omega}_{*}=\dot{\Omega}(\dot{\phi}_{*})$. First we have 

\begin{equation}
\rm \dot{\Omega}_{*}=\frac{\left(\dot{\phi}_{*}+24M_{P}^{2}P_{0}\right)}{2\lambda M_{P}}\,,
\end{equation}
and then 

\begin{equation}
\rm \Omega_{*} = \frac{\sqrt{3}}{6(\lambda^{2}-3)}\left\{2\sqrt{3}(\lambda^{2}-3)\Omega_{0}+24\sqrt{3}\lambda M_{P}P_{0}P_{1} -\log\left[(x_{*}+1)^{\sqrt{3}-\lambda}(x_{*}-1)^{\sqrt{3}+\lambda}\right]\right\} \,,
\end{equation}
where the term $\rm (x_{*}+1)^{\sqrt{3}-\lambda}(x_{*}-1)^{\sqrt{3}+\lambda}$ can be found when inverting Eq.(\ref{number-of-e-fold}); as a result we have that 

\begin{equation}
\rm \Omega(\dot\phi_{*}) = \frac{\sqrt{3}}{6(\lambda^{2}-3)}\left\{2\sqrt{3}(\lambda^{2}-3)(\Omega_{0}-N_{e})+48\sqrt{3}\lambda M_{P}P_{0}P_{1}-\log\left[(x_{end}+1)^{\sqrt{3}-\lambda}(x_{end}-1)^{\sqrt{3}+\lambda}\right]\right\} \,,
\end{equation}
therefore $\rm \Delta_R^2$ in terms of $\rm \dot\phi_{*}$ reads as 

\begin{eqnarray}
\rm \Delta_{\mathcal{R}}^{2} &=& \frac{1}{4\pi^{2}}\frac{1}{24^{2}}\frac{1}{16\lambda^{4}M_{P}^{4}}\frac{\left(\dot\phi_{*}+24M_{P}^{2}P_{0}\right)^{4}}{\dot\phi_{*}^{2}}\times\nonumber\\
&& \times\left\{(x_{end}+1)^{\sqrt{3}-\lambda}(x_{end}-1)^{\sqrt{3}+\lambda}\times\right.\nonumber\\
&& \quad\times\left.\exp\left[-2\sqrt{3}(\lambda^{2}-3)(\Omega_{0}-N_{e})-48\sqrt{3}\lambda M_{P}P_{0}P_{1}\right]\right\}^{\frac{\sqrt{3}}{\lambda^{2}-3}}\,,\label{DeltaR}
\end{eqnarray}
once we fix $\rm \Delta_{\mathcal{R}}^{2}=2.215\times 10^{-9}$ \cite{Planck} one can obtain $\rm \Omega_{0}$ at 50-60 e-folds before inflation ends.

Since we want to contrast the observables parameters with Planck data, we need to evaluate the scalar spectral index $\rm n_{s}$ at horizon crossing, which is defined as  

\begin{equation}
\rm n_{s}-1\equiv \frac{d\ln\Delta_{\mathcal{R}}^{2}}{d\ln k}\Bigr|_{\substack{*}}=\frac{d\ln\Delta_{\mathcal{R}}^{2}}{dN_{e}}\frac{dN_{e}}{d\ln k}\Bigr|_{\substack{*}} \simeq \frac{d\ln\Delta_{\mathcal{R}}^{2}}{dN_{e}}\Bigr|_{\substack{*}} \,,
\end{equation}
where $\rm dN_{e}/d\ln k\simeq 1$, since at horizon crossing ($\rm k=A_{phys *}H_{phys *}$) we have that $\rm \ln k= N_{e}+\ln H_{phys *}$ and $\rm d\ln H_{phys *}/dN_{e}\ll 1$ or $\rm-H'_{phys *}/H_{phys *}^{2}\ll 1$. In order to write down $\rm n_{s}=n_{s}(\dot{\phi}_{*})$; we have 

\begin{equation}
\rm n_{s}-1\simeq\frac{d\ln\Delta_{\mathcal{R}}^{2}}{dN_{e}}\Bigr|_{\substack{*}}=\frac{1}{H_{phys *}\Delta_{\mathcal{R}}^{2}}\frac{d\Delta_{\mathcal{R}}^{2}}{dt_{phys}}\Bigr|_{\substack{*}}=\frac{1}{ \dot\Omega_{*}\Delta_{\mathcal{R}}^{2}}\frac{d\Delta_{\mathcal{R}}^{2}}{dt}\Bigr|_{\substack{*}}\,,
\end{equation}
then the time derivative of the scalar amplitude is

\begin{equation}
\rm \frac{d\Delta_{\mathcal{R}}^{2}}{dt}\Bigr|_{\substack{*}}=2\Delta_{\mathcal{R}}^{2}\left[2\frac{\ddot\Omega_{*}}{ \dot\Omega_{*}}-\frac{\ddot\phi_{*}}{\dot\phi_{*}}-3 \dot\Omega_{*}\right]\,,
\end{equation}
thus the spectral index is

\begin{equation}
\rm n_{s}\simeq -5+4\frac{\ddot\Omega_{*}}{ \dot\Omega_{*}^{2}}-2\frac{\ddot\phi_{*}}{\dot\phi_{*} \dot\Omega_{*}}\,. 
\end{equation}

Once again we need to have a relation of $\rm \ddot{\Omega}_{*}=\ddot{\Omega}(\dot{\phi}_{*})$ and $\rm \ddot{\phi}_{*}=\ddot{\phi}(\dot{\phi}_{*})$. First we have that $\rm\ddot{\Omega}=\ddot{\phi}/(2\lambda M_{P})$, so we need only one relation $\rm \ddot{\phi}_{*}=\ddot{\phi}(\dot{\phi}_{*})$; hence

\begin{equation}
\rm \ddot{\phi}_*=-\frac{1}{2\lambda M_{P}}\left[(\lambda^{2}-3)\dot{\phi}_*^{2}-12^{2}\left(\dot{\phi}_*+12M_{P}^{2}P_{0}\right)M_{P}^{2}P_{0}\right]\,.
\end{equation}
Now we can finally compute the spectral index in terms of the desired parameter, thus 

\begin{equation}\label{ns}
\rm n_{s}\simeq -5-\frac{2(\dot\phi_{*}-24M_{P}^{2}P_{0})}{\dot\phi_{*}(\dot\phi_{*}+24M_{P}^{2}P_{0})^{2}}\left[(\lambda^{2}-3)\dot\phi_{*}^{2}-12^{2}\left(\dot\phi_{*}+12M_{P}^{2}P_{0}\right)M_{P}^{2}P_{0}\right]\,.
\end{equation}

The next observational parameter is the tensor-to-scalar ratio $\rm r$, which is

\begin{equation}
\rm r=\frac{\Delta_{t}^{2}}{\Delta_{\mathcal{R}}^{2}}\,,
\end{equation}
where $\rm \Delta_{t}^{2}=2H_{phys *}^{2}/(\pi^{2}M_{P}^{2})$ is the tensor power spectrum. Then 

\begin{equation}
\rm \Delta_{t}^{2}=\frac{2H_{phys}^{2}}{\pi^{2}M_{P}^{2}}=\frac{2}{\pi^{2}}\frac{\dot{\Omega}_{*}^{2}}{N_{*}^{2}M_{P}^{2}}\,,
\end{equation}
hence

\begin{equation}\label{r}
\rm r = \frac{2^{5}\lambda^{2}\dot\phi_{*}^{2}}{\left(\dot\phi_{*}+24M_{P}^{2}P_{0}\right)^{2}}  \,.
\end{equation}

Finally, combining Eqs.(\ref{ns},\ref{r}) we obtain

\begin{equation}\label{ns-r}
\rm n_{s}=1-\frac{r}{8}+\left[6+\frac{\sqrt{2}(r-96)\lambda}{4\sqrt{r}}\right] \,.
\end{equation}

Eq.(\ref{ns-r}) turns out to be a very interesting result since in the slow-roll approximation $\rm n_{s}$ and $\rm r$ are related as $\rm n_{s}=1-r/8$, which means that the exact solutions of this model yield a general relation between the scalar spectral index and the tensor-to-scalar ratio that includes $\rm \lambda$, which parametrises the slope of the potential. 

Furthermore, as mentioned before, inflation happens in a period such that $\rm A''_{phys}>0$, then $\rm H'_{phys}+H_{phys}^{2}>0$; where in our gauge or proper time, inflation occurs only when $\rm\ddot{\Omega}-2\dot{\Omega}^{2}>0$. Taking the time derivatives to Eq.(\ref{Omegat}) yields

\begin{equation}\label{condition-inflation}
\rm(\lambda^{2}-1)\coth^{2}\left[12\sqrt{3}M_{P}P_{0}(t-P_{1})\right]+\frac{4\sqrt{3}\lambda}{3}\coth\left[12\sqrt{3}M_{P}P_{0}(t-P_{1})\right]+\left(\frac{9-\lambda^{2}}{3}\right)>0\,.
\end{equation}
We have that for $\rm -1<\lambda<1$ the condition that must be satisfied is

\begin{equation}\label{condition-inflation-lambda-small}
\rm\xi_{-}<\coth\left[12\sqrt{3}M_{P}P_{0}(t-P_{1})\right]<\xi_{+}\,,\qquad \xi_{\pm}=\frac{\sqrt{3}\left[-2\lambda\pm (\lambda^2-3) \right]}{3(\lambda^{2}-1)} \,.
\end{equation}

Above condition, Eq.(\ref{condition-inflation-lambda-small}), must be fulfilled during the whole dynamics. Once again, it is convenient to parametrise the evolution with $\dot{\phi}$: we substitute Eq.(\ref{t1}) into Eq.(\ref{condition-inflation-lambda-small}) having

\begin{equation}
\rm\xi_{-}<\frac{\lambda^{2}-3}{24\sqrt{3}\lambda M_{P}^{2}P_{0}}\dot{\phi}-\frac{\sqrt{3}}{\lambda}<\xi_{+}\,,
\end{equation}
after giving a massage to above equation, we have 

\begin{equation}\label{condition-inflation-dotphi-lambda-small}
\rm-\frac{24P_{0}}{\lambda+1}<\frac{\dot{\phi}}{M_{P}^{2}}<\frac{24P_{0}}{\lambda-1}\,.
\end{equation}

This is a very important outcome since we can relate previous results. For instance, the maximum limit of Eq.(\ref{condition-inflation-dotphi-lambda-small}) corresponds to the exact expression of Eq.(\ref{dotphiend}) (with the $\rm \dot{\phi}_{end +}$ root), which precisely indicates when inflation ends regarding any value of $\lambda$. Moreover, having the minimum value $\dot{\phi}$ we can compute the number of e-folds. Following the same notation; substituting the maximum value as $\rm x_{end}(\dot{\phi}_{end}=24P0/(\lambda-1))$ and the minimum as $\rm x_{*}(\dot{\phi}_{*}=-24P0/(\lambda+1))$ in Eq.(\ref{x-dotphi}) such as

\begin{equation}
\rm x_{end}=\frac{\sqrt{3}}{3\lambda}\left(\frac{\lambda^{2}-3}{\lambda-1}\right)-\frac{\sqrt{3}}{\lambda} \,,\qquad x_{*}=-\frac{\sqrt{3}}{3\lambda}\left(\frac{\lambda^{2}-3}{\lambda+1}\right)-\frac{\sqrt{3}}{\lambda}\,, 
\end{equation}
then substituting above equations in Eq.(\ref{number-of-e-fold}), we have 

\begin{equation}\label{Ne-approx-lambda-small}
\rm N_{e} = \frac{\sqrt{3}}{6(\lambda^{2}-3)}\left\{24\sqrt{3}\lambda M_{P}P_{0}P_{1} +\sqrt{3}\log\left[\frac{(\lambda-1)^{2}}{(\lambda+1)^{2}}\right]-\lambda\log[7-4\sqrt{3}]\right\} \,. 
\end{equation}

With this Eq.(\ref{Ne-approx-lambda-small}) one can solve the number of e-folds without the dependency of $\rm\dot{\phi}_{*}$. Besides, with this choice of $\dot{\phi}_{*}$ we find that the tensor-to-scalar ratio is exactly $\rm r=32$. This result implies that $\rm r$ is very large by contrasting to the latest Plank survey 2018 \cite{Planck}, where $\rm r\lesssim 0.064$.

In addition we have that for $\rm 1<\lambda<\sqrt{3}$ and $\rm -\sqrt{3}<\lambda<-1$ the conditions that must be satisfied are 

\begin{equation}\label{condition-inflation-lambda-big}
\rm  \coth\left[12\sqrt{3}M_{P}P_{0}(t-P_{1})\right]<\xi_{-}  \qquad\text{and}\qquad \coth\left[12\sqrt{3}M_{P}P_{0}(t-P_{1})\right]>\xi_{+}\,,
\end{equation}
then performing the same previous procedure we have that 

\begin{equation}\label{condition-inflation-dotphi-lambda-big}
\rm \frac{\dot{\phi}}{M_{P}^{2}}<-\frac{24P_{0}}{\lambda+1} \qquad\text{and}\qquad  \frac{\dot{\phi}}{M_{P}^{2}}>\frac{24P_{0}}{\lambda-1}\,.
\end{equation}

Hence Eqs.(\ref{condition-inflation-dotphi-lambda-small},\,\ref{condition-inflation-dotphi-lambda-big}) are the constraints that must be satisfied in order to have an accelerated expansion. The two distinct domains for $\rm \lambda$, result in two different parameter spaces of $\rm\{P_{0},P_{1},\lambda,\dot{\phi}_{*}\}$, in our analysis we restricted this parameters to those regions that are the most relevant in regards to the observable functions and at the same time ensuring that the inflationary constrictions are satisfied. Besides, given that negative values of $\rm\lambda$ correspond to a mirror-like parameter space with an equivalent outcome, thus we only report the region where $\rm\lambda>0$. Therefore for $\rm 0<\lambda<1$: $\rm P_{0}\,\epsilon\,[-18,0)$ and $\rm P_{1}M_{P}\,\epsilon\,[13,30)$; and for $\rm 1<\lambda<\sqrt{3}$: $\rm P_{0}\,\epsilon\,[-2,0)$ and $\rm P_{1}M_{P}\,\epsilon\,[0,0.5)$. Where in both cases $\rm\dot{\phi}_{*}$ must satisfy Eqs.(\ref{condition-inflation-dotphi-lambda-small},\,\ref{condition-inflation-dotphi-lambda-big}). 

\newpage

\begin{figure}[ht!]
\begin{center}
\captionsetup{width=0.9\textwidth}
\includegraphics[scale=0.5]{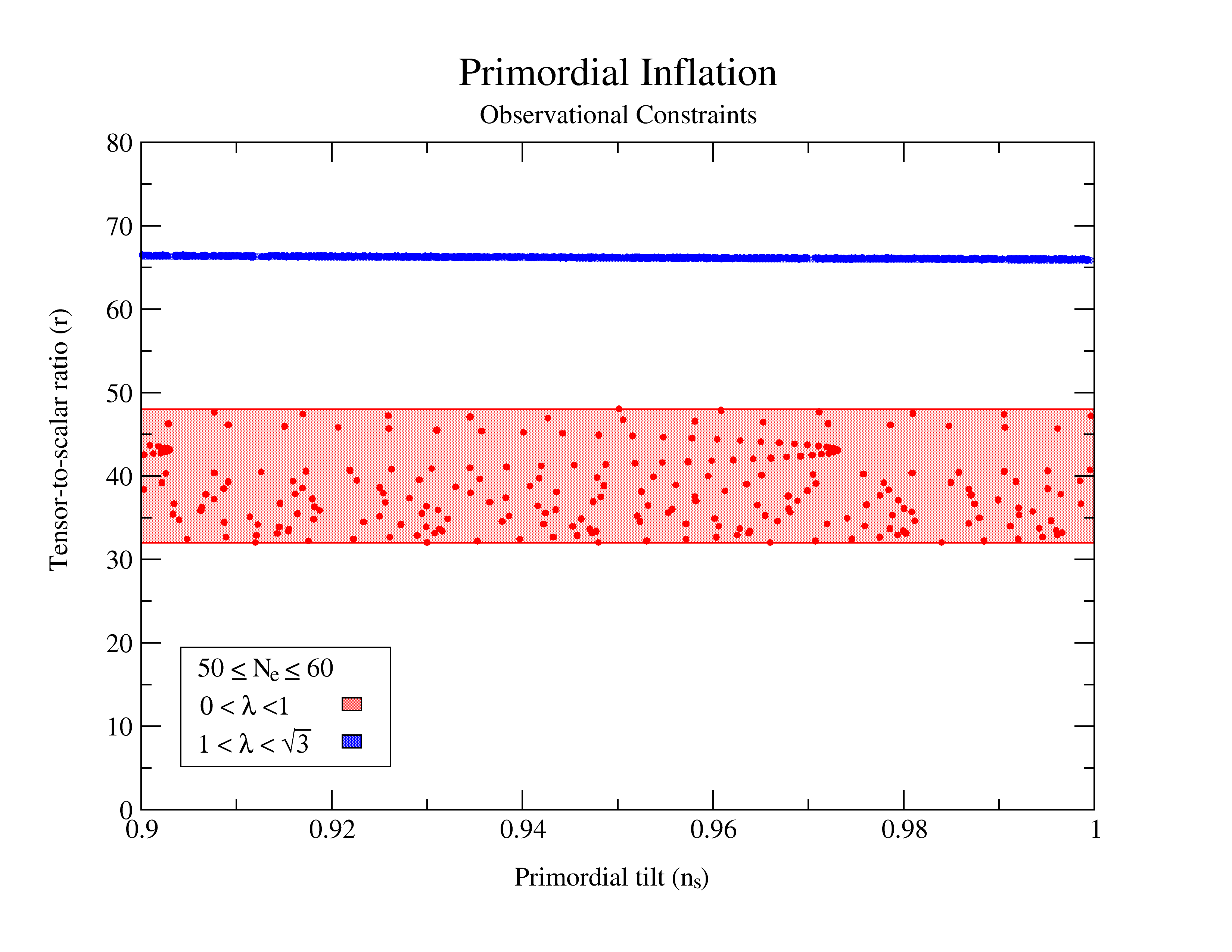}
\caption{Observational predictions with an exponential potential for 50-60 e-folds of inflation. The plot shows two distinct regions, when fixing the primordial tilt $\rm 0.9\leq n_{s}\leq 1$, for the tensor-to-scalar ratio by considering two different domains of $\rm \lambda$. The red contour represents the computed values when $\rm 0<\lambda<1$, where $\rm 32\leq r\lesssim 48$; on the other hand the blue region corresponds to the set of parameters when $\rm 1<\lambda<\sqrt{3}$, having that $\rm 65.45\lesssim r\lesssim 66.85$.}
\label{graf1}
\end{center}
\end{figure}

In Figure \ref{graf1} we present the results when evaluating numerically the observables $\rm n_{s}$ and $\rm r$ at 50-60 e-folds before inflation ends. By fixing the primordial tilt at $\rm 0.9\leq n_{s}\leq 1$ and considering the two different domains of $\rm\lambda$, this yields two well defined regions in the plot for the tensor-to-scalar ratio. The red contour represents the computed values when $\rm 0<\lambda<1$, where $\rm 32\leq r\lesssim 48$; on the other hand the blue region corresponds to the set of parameters when $\rm 1<\lambda<\sqrt{3}$, having that $\rm 65.45\lesssim r\lesssim 66.85$. Indeed, since the contribution of the tensor fluctuations in the tensor-to-scalar ratio is rather small \cite{Planck}, these results picture a disappointing outcome in regards to the phenomenological aspect of this inflationary scenario. Thus, once and for all ruling out this potential when studying the primordial inflation. However, such potential could be relevant in the description of the late time acceleration of the universe in terms of quintessence \cite{Cicciarella}.     

\section{Conclusions\label{conclusions}}

The Quantum approach with a WKB-like ansatz for the wave function, was employed in a Bohmian formalism \cite{bohm}, where the proposal \cite{wssa} was followed in order to find a family of canonical potentials. For the first non trivial case to model inflation, we selected an exponential potential of the form $\rm V=V_{0}e^{-\frac{\lambda\phi}{M_{p}}}$. With this concrete shape of the potential we computed the Hamiltonian's equations of motion using a particular gauge; having thus the exact set of classical solutions of the relevant dynamical parameters. We found that $\rm V_{0}>0$ only when $\rm \lambda^{2}<3$, thus restricting the value of $\rm\lambda$ purely from dynamics. We computed the observable constraints: $\rm\Delta_{\mathcal{R}}^{2}$, $\rm n_{s}$ and $\rm r$, in our proper time; and we were able to evaluate them when the relevant cosmic microwave background (CMB) modes become superhorizon at 50-60 e-folds before inflation ends. We constrained the parameter space: $\rm\{P_{0},P_{1},\lambda,\dot{\phi}_{*}\}$, in order to have an accelerated expansion, following a very restrict set of conditions. To our knowledge, for this particular model of inflation, the observables have not been rigorously evaluated at horizon crossing, since it was believed that this scenario exhibited eternal acceleration when $\rm\lambda<1$ \cite{russo}. Hence, we show that inflation indeed ends regarding any value of $\lambda^{2}<3$. However, the observable parameters present a very discouraging behaviour; for instance by fixing the scalar spectral index within the observational window ($\rm 0.9\leq n_{s}\leq 1$), we found that the tensor-to-scalar ratio is $\rm r\geq 32$, and by contrasting it with $\rm r\lesssim 0.064$ (Planck \cite{Planck}), there is rather large discrepancy with the latest Plank 2018 data. Even though the model is not as phenomenological fitting as expected, the employed method exhibits a remarkable simplicity with rather interesting applications in the near future, perhaps it would require more considerations or further refinement; nevertheless, more potentials or specific models could be analyzed under such procedure.

\acknowledgments{ \noindent
RHJ acknowledges CONACyT for financial support. This work was partially supported by CONACYT  167335, 179881 grants. PROMEP grants UGTO-CA-3. This work is part of the collaboration within the Instituto Avanzado de Cosmolog\'{\i}a and Red PROMEP: Gravitation and
Mathematical Physics under project {\it Quantum aspects of gravity
in cosmological models, phenomenology and geometry of space-time}.
Many calculations where done by the programming language FORTRAN, Symbolic Program REDUCE 3.8. and Wolfram Mathematica 10.0}


\begin{thebibliography}{99}
\bibitem[Alan H. Guth, 1981]{guth1981} 
A.~H.~Guth
     {\it Inflationary universe: A possible solution to the horizon and flatness problem,}
     \emph{Phys. Rev. D}
     {\bf 23}, 347 (1981).
\bibitem[A. D. H. Linde, 1982]{linde1982} 
A.~D.~Linde
     {\it A new inflationary universe scenario: A possible solution of the horizon, flatness, homogeneity, isotropy and primordial monopole problems,}
     \emph{Phys. Lett. B}
     {\bf 108}, 389-193 (1982). 
\bibitem[J. D. Barrow \& M. S. Turner, 1981]{turner1981} 
J.~D.~Barrow and M.~S.~Turner
     {\it Inflation in the Universe,}
     \emph{Nature}
     {\bf 292}, 35-38 (1981) [doi:10.1038/292035a0]. 
\bibitem[Alexei A. Starobinsky, 1980]{starobinsky1980} 
A.~A.~Starobinsky 
	{\it A new type of isotropic cosmological models without singularity,}
	\emph{Phys. Lett. B}
	{\bf 91}, 99 (1980).   
	
\bibitem{Starobinsky:1979ty} 
A.~A.~Starobinsky,
{\it Spectrum of relict gravitational radiation and the early state of the universe,
JETP Lett.}\ {\bf 30}, 682 (1979)
[Pisma Zh.\ Eksp.\ Teor.\ Fiz.\  {\bf 30}, 719 (1979)].
	
\bibitem{Mukhanov:1981xt} 
V.~F.~Mukhanov and G.~V.~Chibisov,
{\it Quantum Fluctuations and a Nonsingular Universe,
JETP Lett.}\  {\bf 33}, 532 (1981)
[Pisma Zh.\ Eksp.\ Teor.\ Fiz.\  {\bf 33}, 549 (1981)].
	
\bibitem[H. Kodama \& M. Sasaki, 1984]{kodama} 
H.~Kodama and M.~Sasaki
	{\it Cosmological Perturbation Theory,}
	\emph{Progress of Theoretical Physics Supplement} 
	{\bf 78}, 1-166 (1984) [doi.org/10.1143/PTPS.78.1]	
	
\bibitem{bassett} 
B.~A.~Bassett, S.~Tsujikawa and D.~Wands, 
    \emph{Rev. Mod. Phys.} 
    {\bf{78}}, 537 (2006) [arXiv:0507632].	

\bibitem{Planck}
P.~A.~R. Ade et al. (Planck Collaboration), {\it Planck 2018 results. X. Constraints on inflation},
[arXiv:1807.06211].
 	
\bibitem[John D. Barrow, 1985]{barrow} 
J.~D.~Barrow
	{\it Slow-roll inflation in scalar-tensor theories,}
	\emph{Phys. Rev. D}
	{\bf 51}, 2729 (1995). 	
\bibitem[A. R. Liddle \& Scherrer, 1998]{andrew1998b}    
A.~R.~Liddle and R.~J.~Scherrer
    {\it Classification of scalar field potential with cosmological scaling solutions,}
  	\emph{Phys. Rev. D}
  	{\bf 59}, 023509 (1998)~[doi.org/10.1103/PhysRevD.59.023509].
  	
\bibitem[Ferreira \& Joyce, 1998]{ferreira2}  
P.~G.~Ferreira and M.~Joyce
    {\it Cosmology with a primordial scaling field,}
    \emph{Phys. Rev. D} {\bf 58}, 023503 (1998)~[doi.org/10.1103/PhysRevD.58.023503].
\bibitem{copeland1} 
E.~J.~Copeland, A.~Liddle and D.~Wands
     {\it Exponential potentials and cosmological scaling solutions,}
     \emph{Phys. Rev. D}
     {\bf 57} 4686, (1998)~[doi.org/10.1103/PhysRevD.57.4686].
\bibitem{copeland2} 
E.~J.~Copeland, T.~Barreiro and N.~J.~Nunes
     {\it Quintessence arising from exponential potentials,}
     \emph{Phys. Rev. D}
     {\bf 61} 127301, (2000)~[doi.org/10.1103/PhysRevD.61.127301].         
\bibitem[Gianluca Calcagni \& Andrew R. Liddle, 2007]{andrew2007} 
G.~Calcagni and A.~R.~Liddle
     {\it Stability of multifield cosmological solutions,}
     \emph{Phys. Rev. D}
      {\bf 77} 023522,(2008)~[doi.org/10.1103/PhysRevD.77.023522].
\bibitem[D. S\'{a}ez-G\'{o}mez, 2008]{gomez} 
D.~S\'{a}ez-G\'{o}mez
     {\it Scalar-Tensor theories and current Cosmology,}
     \emph{Problems of Modern Cosmology}
     (2008) [arXiv:0812.1980].
\bibitem[M. Capone, C. Rubano, P. Scudellaro, 2006]{capone} 
M.~Capone, C.~Rubano and P.~Scudellaro
     {\it Slow rolling, inflation and quintessence,}
     \emph{Europhys.Lett}
     {\bf 73} 149-155, (2006) [arXiv:astro-ph/0607556].

\bibitem[Kolb \& Turner, 1998]{kolb} 
E.~W.~Kolb and M.~S.~Turner,
    {\it The Early Universe,}
    (Addison-Wesley publishing co., Illinois, 1998).


\bibitem{Lucchin:1984yf} 
F.~Lucchin and S.~Matarrese,
{\it Power Law Inflation,
Phys.\ Rev.\ D} {\bf 32}, 1316 (1985)~[doi:10.1103/PhysRevD.32.1316].

\bibitem{Salopek:1990jq} 
D.~S.~Salopek and J.~R.~Bond,
{\it Nonlinear evolution of long wavelength metric fluctuations in inflationary models,
Phys.\ Rev.\ D} {\bf 42}, 3936 (1990)~[doi:10.1103/PhysRevD.42.3936].

\bibitem{ratra1} 
B.~Ratra, 
{\it Quantum mechanics of exponential-potential inflation,}
\emph{Phys. Rev. D} {\bf 40} 3939, (1989)~[doi.org/10.1103/PhysRevD.40.3939].

\bibitem{ratra2} 
B.~Ratra, 
{\it Inflation in an exponential-potential scalar field model,}
\emph{Phys. Rev. D} {\bf 45} 1913, (1992)~[doi.org/10.1103/PhysRevD.45.1913].

\bibitem{russo}
J.~G.~Russo,
{\it Exact solution of scalar field cosmology with exponential potentials and transient acceleration,}
\emph{Phys. Lett. B}
{\bf 600} 185-190, (2004)~[doi.org/10.1016/j.physletb.2004.09.007].

\bibitem{Andrianov:2011fg} 
A.~A.~Andrianov, F.~Cannata and A.~Y.~Kamenshchik,
{\it General solution of scalar field cosmology with a (piecewise) exponential potential,}
JCAP {\bf 1110}, 004 (2011)
[arXiv:1105.4515].

\bibitem{Piedipalumbo:2011bj} 
E.~Piedipalumbo, P.~Scudellaro, G.~Esposito and C.~Rubano,
{\it On quintessential cosmological models and exponential potentials,}
Gen.\ Rel.\ Grav.\  {\bf 44}, 2611 (2012)
[arXiv:1112.0502].

\bibitem{Fre:2013vza} 
P.~Fr\'{e}, A.~Sagnotti and A.~S.~Sorin,
{\it Integrable Scalar Cosmologies I. Foundations and links with String Theory,}
Nucl.\ Phys.\ B {\bf 877}, 1028 (2013)
[arXiv:1307.1910].

\bibitem{obre} 
O.~Obreg\'{o}n, J.~J.~Rosales, J.~Socorro and V.~I.~Tkach,
{\it Supersymmetry breaking a normalizable wavefunction for the FRW (k=0) cosmological model,}
 \emph{Classical and quantum gravity}
{\bf 16} 2861-2870, (1999).


\bibitem{sodo} 
J.~Socorro and M.~D'oleire,
{\it Inflation from supersymmetric  quantum cosmology,}
\emph{Phys. Rev. D}
{\bf 82}(4) 044008, (2010)~[doi.org/10.1103/PhysRevD.82.044008].

\bibitem{ssw} 
J.~Socorro, M.~Sabido and W.~Ram\'irez and M.~G.~Ag\"uero,
{\it Inflaci\'on cosmol\'ogica vista desde la mec\'anica cu\'antica supersim\'etrica}, Ed. Notabilis Scientia (2013).

\bibitem{sdp} 
J.~Socorro, M.~D'oleire and L.~O.~Pimentel,
{\it Time-varying cosmological term,
J.\ Phys.\ Conf.\ Ser. }  {\bf 654}, no. 1, 012007 (2015)~[doi:10.1088/1742-6596/654/1/012007].

\bibitem[Guzm\'an et al., 2007]{wssa} 
W.~Guzm\'an, M.~Sabido, J.~Socorro and L.~A.~Ure\~na-L\'opez,
{\it Scalar potentials out of canonical quantum cosmology,}
\emph{Int. J. Mod.  Phys. D } {\bf 16} (4), 641-653 (2007).
    
\bibitem{nuevo} 
J.~Socorro and O.~E.~N\'u\~{n}ez, 
{\it Scalar potentials with multi-scalar fields from quantum cosmology an supersymetric quantum mechanics},
{\emph Eur. Phys. Journal Plus} {\bf 132}: 168 (2017) [arXiv:1702.00478]. 

\bibitem[Gibbons \& Gishchuk, 1989]{Gibbons} 
G.~W.~Gibbons and L.~P.~Grishchuk,
    {\it Nucl. Phys. B}
    {\bf 313}, 736 (1989).
    
\bibitem[Zhi, 1987]{Zhi}  
L.~Z.~Fang and R.~Ruffini,
    {\it Quantum Cosmology,  Advances Series in Astrophysics and
    Cosmology Vol. 3,}
    (World Scientific, Singapore, 1987).
    
\bibitem[Hartle \& Hawking, 1983]{HH} 
J.~B.~Hartle and S.~W.~Hawking,
    \emph{Phys. Rev. D},
    {\bf 28}, 2960 (1983).
    
\bibitem[Hawking, 1984]{H} 
S.~W.~Hawking,
    \emph{ Nucl. Phys. B}
    {\bf 239}, 257 (1984).
    
\bibitem[Bohm, 1952]{bohm} 
D.~Bohm,
    {\it Suggested interpretation of the quantum theory in terms of  "Hidden" variables I,}
    \emph{Phys. Rev.}
    {\bf 85} (2), 166 (1952). 

\bibitem{Cicciarella} 
F.~Cicciarella and M.~Pieroni, 
    {\it Universality for quintessence}, 
    JCAP {\bf{08}} (2017) 010 [arXiv:1611.10074].

\end{thebibliography}
\end{document}